\documentclass{llncs}

\usepackage{hyperref}
 \usepackage{rotating}

\usepackage{subfigure}
\usepackage{epsfig}
\usepackage{psfrag}
\usepackage{graphicx}

\usepackage{float}
\usepackage{amssymb}

\usepackage{amsmath}

\usepackage{algorithm,myalgorithmic}
\renewcommand{\algorithmicrequire}{\textbf{Input(s):}}
\renewcommand{\algorithmicensure}{\textbf{Output(s):}}

\newcommand{\comment}[1]{ }

\spnewtheorem{obs}[theorem]{Observation}{\bfseries}{\itshape}

\newcommand{\Oh}{{\mathcal{O}}}

\renewcommand{\int}{\operatorname{int}}       

\newcommand{\NP}{\mbox{$\mathcal{NP}$}}

\newcommand{\true}{$\cal T$}
\newcommand{\ttrue}{{\bf \sf true}}

\newcommand{\ffalse}{{\bf \sf false}}

\newcommand{\scp}{{\it scp}}

\newcommand{\rut}[1]{$\Oh^*(#1)$}

\newcommand{\rt}{2^{\frac{K}{6.1489}}}
\newcommand{\rtt}{2^{\frac{K}{6.2158}}}

\renewcommand{\subsubsection}[1]{\smallskip\noindent{\bf #1}} 

\begin{document}
\title{A New Upper Bound for Max-2-SAT:\\ A Graph-Theoretic Approach}

\author{
  Daniel Raible   \&  Henning Fernau 
    }
\institute{%
University of
Trier, FB 4---Abteilung Informatik, 
54286 Trier,
Germany
\text{\{raible,fernau\}@informatik.uni-trier.de}}
\maketitle
\begin{abstract}
In {\sc MaxSat}, we ask for an assignment which satisfies the maximum number of clauses for a boolean formula in CNF. We present an algorithm yielding a run time upper bound of $\Oh^*(\rtt )$ for {\sc Max-2-Sat} (each clause contains at most 2 literals), where $K$ is the number of clauses. The run time has been achieved by using heuristic priorities on the choice of the variable on which we branch. The implementation of these heuristic priorities is rather simple, though they have a significant effect on the run time. Also  the analysis uses a non-standard measure. 
\end{abstract}

\section{Introduction}
\subsubsection{Our Problem}. 
{\sc MaxSat} is an optimization version of the well-known decision problem {\sc SAT}: given a boolean formula in CNF, we ask for an assignment which satisfies the maximum number of clauses. 
The applications for {\sc MaxSat} range over such fields as  combinatorial optimization, artificial intelligence and database-systems as mentioned in \cite{KojKul2006}.  We put our focus on {\sc Max-2-Sat}, where every formula is constrained to have at most two literals per clause,
to which problems as  {\sc Maximum Cut} 
and {\sc Maximum Independent Set} are reducible. 
Therefore {\sc Max-2-Sat} is \NP-complete.

\subsubsection{Results So Far}. The best published upper bound of $\Oh^*(2^{\frac{K}{5.88}})$  has been achieved by 
 Kulikov and Kutzov in \cite {KulKut2007} consuming only polynomial space. They build up their algorithm on the one of  Kojevnikov and Kulikov~\cite{KojKul2006} who were the first who used a non-standard measure yielding a run time of \rut{2^{\frac{K}{5.5}}}.
If we measure the complexity in the number $n$ of variables the current fastest algorithm is the one of R. Williams \cite{Wil2005} having run time $\Oh^*(2^{\frac{\omega}{3}n})$, where $\omega <2.376$ is the matrix-multiplication exponent. A drawback of this algorithm is its requirement of exponential space. Scott and Sorkin \cite{ScoSor2007} presented a $\Oh^*(2^{1 - \frac{1}{d+1}n})$-algorithm consuming polynomial space, where $d$ is the average degree of the variable graph.
{\sc Max-2-Sat} has also been studied with respect to approximation \cite{Hof2003,LewLivZwi2002} and parameterized algorithms \cite{Graetal03a,GraNie2000}.

\subsubsection{Our Results}. The major result we present is an algorithm solving {\sc Max-2-Sat} in time $\Oh^*(\rtt)$. Basically it is a refinement of the algorithm in \cite{KojKul2006}, which also in turn builds up on the results of \cite{Graetal03a}. The run time improvement is twofold. In \cite{KojKul2006} an upper bound of $\Oh^*( 1.1225^n)$ is obtained if the variable graph is cubic. Here $n$ denotes the number of variables. We could improve this to $\Oh^*(1.11199^n)$ by a more accurate analysis. Secondly, in the case where the maximum degree of the variable graph is four, we choose a variable for branching according to some heuristic priorities. These two improvements already give a run time of \rut{\rt}. Moreover we like to point out that these heuristic priorities can be implemented such that they only consume $\Oh(n)$ time. The authors of \cite{KulKut2007} improve the algorithm of \cite{KojKul2006} by having a new branching strategy when the variable graph has maximum degree five.  Now combining our improvements with the ones from \cite{KulKut2007} gives the claimed run time. 

\subsubsection{Basic Definitions and Terminology}.
Let $V(F)$ be the set of variables of a given boolean formula $F$.
For $v \in V(F)$ by $\bar{v}$ we denote the negation of $v$. If $v$ is set, then it will be assigned the values \ttrue\   or \ffalse.  By the word \emph{literal}, we refer to a variable or its negation.
 A \emph{clause} is a disjunction of literals.
We consider formulas in \emph{conjunctive normal form (CNF)}, that is a conjunction of clauses.  We allow only 1- and 2-clauses, i.e., clauses with at most two literals.
The weight of $v$, written $\#_2(v)$, refers to the number of 2-clauses in which $v$ or $\bar{v}$ occurs. For a set $U\subseteq V(F)$ we define $\#_2(U):=\sum_{u \in U} \#_2(u)$. If $v$ or $\bar{v}$ occurs in some clause $C$ we write $v \in C$.
 A set $A$ of literals is called \emph{assignment} if for every $v \in A$ it holds that $\bar{v} \not \in A$. Loosely speaking if $l \in A$ for a literal $l$, than $l$ receives the value \ttrue. We allow the formula to contain truth-clauses of the form \{\true\} that are always satisfied.  Furthermore, we consider a {\sc Max-2-Sat} instance as multiset of clauses.
 A $x \in V(F)$ is a \emph{neighbor} of $v$, written $x \in N(v)$, if they occur in a common 2-clause. Let $N[v]:=N(v)\cup \{v\}$. The \emph{variable graph} $G_{var}(V,E)$ is defined as follows: $V=V(F)$ and $E=\{\{u,v\} \mid u,v \in V(F), u \in N(v)\}$. 
Observe that $G_{var}$ is a undirected multigraph and that it neglects clauses of size one. 
We will not distinguish between the  words ``variable'' and ``vertex''. Every variable in a formula corresponds to a vertex in $G_{var}$ and vice versa. 
By writing $F[v]$, we mean the formula which emerges from $F$ by setting $v$ to \ttrue\  the following way:
First, substitute all clauses containing $v$ by \{\true\}, then
delete all occurrences of $\bar{v}$ from any clause and finally delete all empty clauses from $F$.
$F[\bar{v}]$ is defined analogously: we set $x$ to \ffalse. 
\section{Reduction Rules \& Basic Observations}\label{sec2}
\label{RR}
We state well-known reduction rules from previous work \cite{Graetal03a,KojKul2006}:\\[1ex]
\comment{ 
\begin{description}
\item[{\bf RR-1}] Replace any 2-clause $C$ with $l,\bar{l} \in C$, for a literal $l$, with \{\true\}.
\item[{\bf RR-2}] If for two clauses $C,D$ and a literal $l$ we have $C \setminus \{l\} = D \setminus \{\bar{l}\}$, then substitute $C$ and $D$ by $C \setminus \{l\}$ and \{\true\}.
\item[{\bf RR-3}] A literal $l$ occurring only positively (negatively, resp.) is set to \ttrue\ (\ffalse).
\item[{\bf RR-4}] If $\bar{l}$ does not occur in more 2-clauses than $l$ in 1-clauses, such that $l$ is a literal, then set $l$ to \ttrue.
\item[{\bf RR-5}] Let $x_1$ and $x_2$ be two variables, such that $x_1$ appears at most once in  another clause without $x_2$. In this case, we call $x_2$ the \emph{companion} of $x_1$. {\bf RR-3} or {\bf RR-4}  will  set $x_1$ in $F[x_2]$ to $\alpha$ and in $F[\bar{x}_2]$ to $\beta$, where $\alpha,\beta \in \{\ttrue, \ffalse\}$. Depending on $\alpha$ and $\beta$, the following actions will be carried out:
\begin{itemize}
\item If $\alpha=\ffalse$, $\beta=\ffalse$, set $x_1$ to \ffalse.
\item If $\alpha=\ttrue$, $\beta=\ttrue$, set $x_1$ to \ttrue.
\item If $\alpha=\ttrue$, $\beta=\ffalse$, substitute every occurrence of $x_1$ by $x_2$.
\item If $\alpha=\ffalse$, $\beta=\ttrue$, substitute every occurrence of $x_1$ by $\bar{x}_2$.
\end{itemize}
\end{description}
}
{\bf RR-1} Replace any 2-clause $C$ with $l,\bar{l} \in C$, for a literal $l$, with \{\true\}. \\
{\bf RR-2} If for two clauses $C,D$ and a literal $l$ we have $C \setminus \{l\} = D \setminus \{\bar{l}\}$, then substitute $C$ and $D$ by $C \setminus \{l\}$ and \{\true\}.\\
{\bf RR-3} A literal $l$ occurring only positively (negatively, resp.) is set to \ttrue\ (\ffalse).\\
{\bf RR-4} If $\bar{l}$ does not occur in more 2-clauses than $l$ in 1-clauses, such that $l$ is a literal, then set $l$ to \ttrue.\\
{\bf RR-5} Let $x_1$ and $x_2$ be two variables, such that $x_1$ appears at most once in  another clause without $x_2$. In this case, we call $x_2$ the \emph{companion} of $x_1$. {\bf RR-3} or {\bf RR-4}  will  set $x_1$ in $F[x_2]$ to $\alpha$ and in $F[\bar{x}_2]$ to $\beta$, where $\alpha,\beta \in \{\ttrue, \ffalse\}$. Depending on $\alpha$ and $\beta$, the following actions will be carried out:
\begin{description}
\item If $\alpha=\ffalse$, $\beta=\ffalse$, set $x_1$ to \ffalse.
\item If $\alpha=\ttrue$, $\beta=\ttrue$, set $x_1$ to \ttrue.
\item If $\alpha=\ttrue$, $\beta=\ffalse$, substitute every occurrence of $x_1$ by $x_2$.
\item If $\alpha=\ffalse$, $\beta=\ttrue$, substitute every occurrence of $x_1$ by $\bar{x}_2$.
\end{description}
From now on we will only consider reduced formulas $F$. This means that to a given formula $F$ we apply the following procedure: {\bf RR-i} is always applied before {\bf RR-i+1}, each reduction rule is carried out exhaustively and after {\bf RR-5} we start again with {\bf RR-1} if the formula changed. A formula for which this procedure does not apply will be called \emph{reduced}.
Concerning the reduction rules we have the following lemma~\cite{KojKul2006}:
\begin{lemma}\label{lem0}
\begin{enumerate}
\item If $\#_2(v)=1$, then $v$ will  be set. \label{obs1}\label{cor1} \label{lem01}
\item For any $u \in V(F)$ in a reduced formula with $\#_2(u)=3$ we have $|N(v)|=3$.\label{lem03a}
\item If the variables $a$ and $x$ are neighbors and $\#_2(a)=3$, then in at least one of the formulas $F[x]$ and $F[\bar{x}]$, the reduction rules set $a$.\label{lem03b}
\end{enumerate}
\end{lemma}
We need some auxiliary notions:
A sequence of distinct vertices $a_1,v_1,\ldots, v_j, a_2$ \\ ($j \ge 0$) is called \emph{lasso}   if $\#_2(v_i)=2$ for $1 \le i \le j$, $a_1=a_2$, $\#_2(a_1) \ge 3 $ and $G_{var}[a_1,v_1,\ldots, v_j, a_2]$ is a cycle. A \emph{quasi-lasso} is a lasso with the difference that $\#_2(v_j)=3$. A lasso is called \emph{3-lasso} (resp. 4-lasso) if $\#_2(a_1)=3$ ($\#_2(a_1)=4$, resp.). 
\emph{3-quasi-lasso} and \emph{4-quasi-lasso} are defined analogously.

\begin{lemma}\label{obs2}
\begin{enumerate}
\item Let $v,u,z \in V(F)$ be pairwise distinct with $\#_2(v)=3$ such that there are clauses $C_1,C_2,C_3$ with $u,v \in C_1,C_2$ and $v,z \in C_3$. 
Then either $v$ is set or the two common  edges of $u$ and $v$ will be contracted in $G_{var}$.\label{lem02}
\item The reduction rules delete  the variables  $v_1,\ldots, v_j$ of a lasso (quasi-lasso, resp.) and the weight of $a_1$ drops by at least two (one, resp.).\label{lemlem}
\end{enumerate}
\end{lemma}
\begin{proof}
\begin{enumerate}
\item If $v$ is not set it will be substituted by $u$ or $\bar{u}$ due to {\bf RR-5}. The emerging clauses $C_1,C_2$ will be reduced either by {\bf RR-1} or become  1-clauses. Also we have an edge between $u$ and $z$ in $G_{var}$ as now $u,z \in C_3$. 
\item We give the proof by induction on $j$. In the lasso case for $j=0$, there must be a  2-clause $C=\{a_1,\bar{a}_1\}$, which will be deleted by {\bf RR-1}, so that the initial step is shown. So now $j>0$. Then on any $v_i$, $1 \le i \le j$, we can apply {\bf RR-5} with any neighbor as companion, so, w.l.o.g., it is applied to $v_1$ with $a_1$ as companion. {\bf RR-5} either sets $v_1$, then we are done with   Lemma~\ref{lem0}.\ref{cor1}, or $v_1$ will be substituted by $a_1$. By applying {\bf RR-1},  this leads to the lasso $   a_1,v_2,\ldots, v_j, a_2$ in $G_{var}$ and the claim follows by induction. 
In the quasi-lasso case for $j=0$, the arguments from above hold. For $j=1$, item $1.$ is sufficient. For $j>1$, the induction step from above also applies here.\qed
\end{enumerate}
 \end{proof}

\section{The Algorithm}
We set $d_i(F):=|\{x \in V(F) \mid \#_2(x)=i \}|$. 
To measure the run time, we choose a non standard measure approach with the measure $\gamma$ defined as follows:\\[1ex]
$
\mbox{\ \ }\gamma(F)=\sum_{i=3}^{n} \omega_i \cdot d_i(F) \mbox{ with } \omega_3=0.94165   , \: \omega_4 = 1.80315, \: \omega_i = \frac{i}{2} \mbox{ for } i \ge 5.
$\\[1ex]
Clearly, $\gamma(F)$ never exceeds the number of clauses $K$ in the corresponding formula. So, by showing  an upper bound of $c^{\gamma(F)}$ we can infer an upper bound $c^K$.
We set $\Delta_3:=\omega_3$, $\Delta_i:=\omega_i -\omega_{i-1}$ for $i \ge 4$. Concerning the $\omega_i$'s we have  $\Delta_i \ge \Delta_{i+1}$ for $i \ge 3$ and $\omega_4 \ge 2 \cdot \Delta_4$. 
The algorithm presented in this paper proceeds as follows: After applying the  above-mentioned reduction rules exhaustively, it will branch on a variable $v$. That is, we will reduce the problem to the two formulas $F[v]$ and $F[\bar{v}]$. In each of the two branches, we must determine by how much the original formula $F$ will be reduced in terms of $\gamma(F)$. Reduction in $\gamma(F)$ can be due to branching on a variable or to the subsequent application of reduction rules. By an \emph{$(a_1,\ldots,a_\ell)$-branch}, we mean that in the $i$-th branch $\gamma(F)$ is reduced by  at least $a_i$. The \emph{i-th component} of a branch refers to the search tree evolving from the $i$-th branch (i.e., $a_i$). By writing $(\{a_1\}^{i_1},\ldots,\{a_\ell \}^{i_\ell})$-branch we mean a $(a_1^1,\ldots,a_1^{i_1},\ldots,a_\ell^1,\ldots,a_\ell^{i_\ell})$-branch where $a_j^s=a_j$ with $1 \le s \le i_j$. A $(a_1,\ldots, a_\ell)$-branch \emph{dominates} a $(b_1,\ldots ,b_\ell)$-branch if $a_i \ge b_i$ for $1 \le i\le \ell$.

\subsubsection{Heuristic Priorities}
If the maximum degree of $G_{var}$ is four,  variables $v$ with $\#_2(v)=4$ will be called \emph{limited} if there is another variable $u$ appearing with  $v$ in two 2-clauses (i.e., we have two edges between $v$ and $u$ in $G_{var}$). We call such $u,v$ a \emph{limited pair}. Note that also $u$ is limited and that at this point by {\bf RR-5} no two weight 4 variables can appear in more than two clauses together. We call $u_1,\ldots u_\ell$ a \emph{limited sequence} if $\ell \ge 3$ and $u_i,u_{i+1}$ with $1 \le i \le \ell-1$  are limited pairs. A \emph{limited cycle} is a limited sequence with $u_1=u_\ell$.
To obtain an asymptotically fast algorithmic behavior we introduce heuristic priorities  ({\bf HP}), concerning the choice of the variable used for branching. 
\begin{enumerate}
\item Choose any $v$ with $\#_2(v) \ge 7$.
\item Choose any $v$ with $\#_2(v)=6$, preferably with $\#_2(N(v))<36$.
\item Choose any $v$ with $\#_2(v)=5$, preferably with $\#_2(N(v))<25$.
\item Choose any unlimited $v$ with $\#_2(v)=4$ and a limited neighbor.
\item Choose the vertex $u_1$ in a limited sequence or cycle.
\item Pick a limited pair $u_1,u_2$. Let $c \in N(u_1)\setminus \{u_2\} $ with $s(c):=|(N(c) \cap (N(u_1)) \setminus \{c,u_1\})|$ maximal. If $s(c)>1$, then choose the unique vertex in $N(u_1) \setminus \{u_2,c\}$, else choose $u_1$.
\item From $Y:=\{v \in V(F) \mid \#_2(v)=4, \exists z \in N(v):\; \#_2(z)=3 \wedge N(z) \not \subseteq N(v)\}$ choose $v$, preferably such that  $\#_2(N(v))$ is maximal. 
\item Choose any $v$, with $\#_2(v)=4$, preferably with  $\#_2(N(v))<16$.
\item Choose any $v$, with $\#_2(v)=3$, such that there is $a \in N(v)$, which forms a triangle $a,b,c$ and $b,c \not \in N[v]$ (we say $v$ has \emph{pending triangle} $a,b,c$).
\item Choose any $v$, such that we have a $(6\omega_3,8\omega_3)$- or a $(4\omega_3,10 \omega_3)$-branch.
\end{enumerate}
From now on $v$ denotes the variable picked according to {\bf HP}.
\begin{algorithm}
{\bf Procedure:} SolMax2Sat($F$)
\begin{algorithmic}[1]
\STATE Apply SolMax2Sat on every component of $G_{var}$ separately.
\STATE Apply the reduction rules exhaustively to $F$.
\STATE Search exhaustively on any sub-formula being a component of at most 9 variables. 
\IF{$F=$ \{\true\}$\ldots$\{\true\}} 
\STATE {\bf return} $|F|$
\ELSE
\STATE Choose a variable $v$ according to {\bf HP}.
\STATE {\bf return} $\max\{\mbox{SolMax2SAT}(F[v]),\mbox{SolMax2Sat}(F[\bar{v}])\}$.
\ENDIF
\end{algorithmic}
\caption{An algorithm for solving {\sc Max-2-Sat}.}
\label{algo1}
\end{algorithm}
\noindent

\subsubsection{Key Ideas} The main idea is to have some priorities on the choice of a weight 4 variable such that the branching behavior is beneficial. For example limited variables tend to be  unstable in the following sense: If their weight is decreased due to branching they will be reduced due to Lemma~\ref{obs2}.\ref{lem02}. This means we can get an amount of $\omega_4$ instead of $\Delta_4$. In a graph lacking limited vertices we want a variable $v$ with a weight 3 neighbor  $u$ such that $N(u) \not \subseteq N(v)$. In the branch on $v$ where $u$ is set (Lemma~\ref{lem0}.\ref{lem03b}) we can gain some extra reduction (at least $\Delta_4$) from $N(u) \setminus N(v)$. If we fail to find  a variable according to priorities 5-7 we show that  either $v$ as four weight 4 variables and that the graph is 4-regular, or otherwise we have two distinct situations which can be handled quite efficiently.
Further,  the most critical branches are when we have to choose $v$ such that all variables in $N[v]$ have weight $\omega_i$. Then the reduction in $\gamma(F)$ is minimal (i.e., $\omega_i+i\cdot \Delta_i$).
 We analyze this regular case together with its immediate preceding branch. Thereby we prove a better  branching behavior compared to a separate analysis.
In \cite{ScoSor2007} similar ideas were used for {\sc Max-2-CSP}.
We are now ready to present our algorithm, see Alg.~\ref{algo1}. Reaching step 7 we can rely on the fact that $G_{var}$ has at least 10 vertices. We call this the \emph{small component property (\scp)} which is crucial for some cases of the analysis.

\section{The Analysis}\label{analysis}
In this section we investigate the cases when we branch on vertices picked according to  items 1-10 of {\bf HP}. For each item we will derive a branching vector which upper bounds this case in terms of $K$. In the rest of this section we show:
\begin{theorem}
Algorithm~\ref{algo1} has a run time of \rut{\rt}.
\end{theorem}
\subsection{$G_{var}$ has Minimum Degree Four}
\subsubsection{Priority 1}
If $\#_2(v)\ge 7$, we first obtain a reduction of $\omega_7$ because $v$ will be deleted. Secondly, we get an amount of at least $7 \cdot \Delta_7$ as the weights of $v$'s neighbors each drops by at least one and we have $\Delta_i \ge \Delta_{i+1}$. Thus, $\gamma$ is reduced by at least 7 in either of the two branches (i.e., we have a $(\{7\}^2)$-branch).\\
{\it Regular Branches} We call a branch \emph{h-regular} if we branch on a variable $v$ such that for all $u \in N[v]$ we have $\#_2(u)=h$. We will handle those in a separate part. During our considerations a $4$-regular branch will have exactly four neighbors as otherwise this situation is handled by priority 4 of {\bf HP}. The following subsections handle \emph{non-regular} branches, which means that we can find a $u \in N(v)$ with $\#_2(u)< \#_2(v)$. Note that we already handled  $h$-regular branches for $h \ge 7$.

\subsubsection{Priorities 2 and 3}
 Choosing $v \in V(F)$ with $\#_2(v)=6$ there is a $u \in N(v)$ with $\#_2(u)\le 5$ due to non-regularity. Then by deletion of $v$, there is a reduction by $\omega_6$ and another of at least $5\Delta_6 + \Delta_5$, resulting from the dropping weights of the neighbors. Especially, the weight of $u$ must drop by at least $\Delta_5$. This leads to a $(\{ 6.19685\}^2)$-branch.
If $\#_2(v) =  5$, the same observations as in the last choice lead  to a reduction of at least $\omega_5+4\cdot \Delta_5+\Delta_4$. Thus we have a $(\{ 6.1489\}^2)$-branch.

\subsubsection{Priority 4}
Let $u_1 \in N(v)$ be the limited variable. $u_1$ forms a limited pair with some $u_2$. 
 After branching on $v$, the variable $u_1$ has weight at most 3. At this point, $u_1$ appears only with one other variable $z$ in a 2-clause. 
Then, {\bf RR-5} is applicable to $u_1$ with $u_2$ as its companion. According to Lemma~\ref{obs2}.\ref{lem02}, either $u_1$ is set or the two  edges of $u_1$ and $u_2$ will be contracted. In the first case, we receive a total reduction of at least $3\omega_4 + 2 \Delta_4$, in the second of at least $2 \omega_4+4 \Delta_4$.
Thus, a proper estimate is a $(\{2 \omega_4+ 4 \Delta_4\}^2)$-branch, i.e., a $(\{7.0523\}^2)$-branch.

\subsubsection{Priority 5}
If $u_1, \ldots, u_\ell$ is a limited cycle, then $\ell \ge 10$ due to \scp. By {\bf RR-5} this yields a $(10w_4,10w_4)$-branch.
If $u_1,\ldots, u_\ell$ is a limited sequence, then  due to priority 4 the neighbors of $u_1,u_\ell$ lying outside the sequence have weight 3. By {\bf RR-5} the branch on $u_1$ is a $(\{3\omega_4+2\omega_3\}^2)$-branch, i.e, a $(\{7.29275\}^2)$-branch.

\subsubsection{Priority 6}
At this point every limited variable $u_1$ has two neighboring variables $y,z$ with weight 3 and a limited neighbor $u_2$ with the same properties 
(due to priorities 4 and 5). We now examine the local structures arising from this fact and by the values of $|N(y)\setminus N(u_1)|$ and $|N(z) \setminus N(u_1)|$. 

\begin{enumerate}
\item We rule out $|N(y)\setminus N(u_1)|=|N(z)\setminus N(u_1)|=0$ due to \scp.
\item $|N(y)\setminus N(u_1)|=0, |N(z)\setminus N(u_1)|=1$:  Then, $N(y)=\{u_2,z,u_1\},N(u_2)=\{u_1,y,s_1\}$ and $N(z)=\{u_1,y,s_2\}$, see Figure~\ref{prio62}. In this case we branch on $z$ as $s(y)>0$ and $s(y)>s(z)$. Then due to  {\bf RR-5} $y$ and $u_1$ disappear; either by being set or replaced. Thereafter due to {\bf RR-1} and Lemma~\ref{lem0}.\ref{obs1} $u_2$ will be set. Additionally we get an amount of $\min\{2 \Delta_4,\omega_4,\omega_3+\Delta_4\}$ from $s_1,s_2$. This depends on  whether $s_1 \neq s_2$  or $s_1=s_2$ and in the second case on the weight of $s_1$. If $\#_2(s_1)=3$ we get a reduction of $\omega_3+\Delta_4$ due to setting $s_1$. In total we have at least a $(\{2\omega_4+2\omega_3+2\Delta_4\}^2)$-branch. 
Analogous is the case $|N(y)\setminus N(u_1)|=1, |N(z)\setminus N(u_1)|=0$.
\item $|N(y)\setminus N(u_1)|=1, |N(z)\setminus N(u_1)|=1$: Here two possibilities occur:\\
{\it (a)} $N(y)=\{u_1,u_2,s_1\},N(z)=\{u_1,u_2,s_2\}$, $N(u_2)=\{u_1,y,z\}$, see Figure~\ref{prio63a}: Then w.l.o.g., we branch on $z$. Similarly to item $2.$ we obtain a $(\{2\omega_4+2\omega_3+2\Delta_4\}^2)$-branch.\\
{\it (b)} $N(y)=\{u_1,z,s_1\}$, $N(z)=\{u_1,y,s_2\}$, see Figure~\ref{prio63b}: W.l.o.g., we branch on $z$. Basically we get a total reduction of $\omega_4+2\omega_3+2\Delta_4$. That is $2\omega_3$ from $y$ and $z$, $\omega_4$ from $u_1$ and $2\Delta_4$ from $s_2$ and $u_2$. In the branch where $y$ is set (Lemma~\ref{lem0}.\ref{lem03b}) we additionally get $\Delta_4$ from $s_1$ and $\omega_4$ from $u_2$ as it will disappear (Lemma~\ref{lem0}.\ref{lem03a}). This is a $(2\omega_4+2\omega_3+2\Delta_4,\omega_4+2\omega_3+2\Delta_4)$-branch.
\item $|N(y)\setminus N(u_1)|=1,|N(z)\setminus N(u_1)|=2$, see Figure~\ref{prio63c}: We branch on $z$ yielding a $(\{2\omega_4+2\omega_3+2\Delta_4\}^2)$-branch. Analogous is the case $|N(y)\setminus N(u_1)|=2,|N(z)\setminus N(u_1)|=1$.
\item $|N(y)\setminus N(u_1)|=2,|N(z)\setminus N(u_1)|=2$: In this case we chose $u_1$ for branching. 
Essentially we get a reduction of $2\omega_4+2\omega_3$. In the branch setting $z$ we receive an extra amount of $2 \Delta_4$ from $z$'s two neighbors outside $N(u_1)$. Hence we have a $(2\omega_4+2\omega_3+2 \Delta_4,2\omega_4+2\omega_3)$-branch.
\end{enumerate}
We have at least a $(2 \omega_4 + 2 \omega_3 + 2 \Delta_4,  \omega_4 + 2 \omega_3 +2\Delta_4)$-branch, i.e., a $(7.2126,5.40945)$-branch.

\subsubsection{Priority 7}
We need further auxiliary notions:
A \emph{3-path} (\emph{4-path}, resp.) for an unlimited weight 4 vertex $v$ is a sequence of vertices $u_0u_1 \ldots u_l  u_{l+1}$ ($u_0u_1 \ldots u_l$, resp.)  forming a path, such that $1 \le l \le 4$ ($2 \le l \le 4$, resp.), $u_i \in N(v)$ for $1 \le i \le l$, $\#_2(u_i)=3$ for $1 \le i \le l$ ($\#_2(u_i)=3$ for $1 \le i \le l-1,\#_2(u_l)=4$, resp.) and $u_0,u_{l+1} \not\in N(v)$ ($u_0 \not \in N(v)$, resp.).
Due to the absence of limited vertices, every vertex $v$, chosen due to priority 7, must have a 3- or 4-path. 
\begin{description}
\item[3-path]\ 
If $u_0\neq u_{l+1}$  we basically get a reduction of $\omega_4+l\omega_3+(4-l)\Delta_4$. In the branch where $u_1$ is set, $u_2 \ldots u_{l}$ will be also set due to Lemma~\ref{lem0}.\ref{lem01}. Therefore, we gain an extra amount of at least $2\Delta_4$ from $u_0$ and $u_{l+1}$, leading to a $(\omega_4+l\cdot \omega_3 + (6-l)\Delta_4 ,\omega_4 + l \cdot \omega_3 + (4-l)\Delta_4)$-branch.\\
If $u_0 = u_{l+1}$ then  in $F[v]$ and in $F[\bar{v}]$, $u_0u_1 \ldots u_l u_{l+1}$ is a lasso. So by Lemma~\ref{obs2}.\ref{lemlem}, $u_1,\ldots,u_l$ are deleted and the weight of $u_0$ drops by 2. If $\#_2(u_0)=4$ this yields a reduction of $l \cdot \omega_3+\omega_4$. If $\#_2(u_0)=3$ the reduction is $(l+1) \cdot \omega_3$ but then $u_0$ is set. It is not hard to see that this yields a bonus reduction of $\Delta_4$ (see Appendix~\ref{Addarg}). Thus, we have a $(\{ \omega_4 + (l +1) \cdot \omega_3+(5-l)\Delta_4\}^2)$-branch. 
\item[4-path] We get an amount of $\omega_4+(l-1)\omega_3+(5-l)\Delta_4$  by deleting $v$. In the branch where $u_1$ is set we get a bonus of $\Delta_4$ from $u_0$. Further $u_l$ will be deleted completely. Hence we have a $(2\omega_4+(l-1)\omega_3 + (5-l)\Delta_4,\omega_4+(l-1)\omega_3+(5-l)\Delta_4)$-branch. 
\end{description}
The first branch is worst for $l=1$, the second and third for $l=2$ (as $l=1$ is impossible). Thus, we have   $(\{7.2126\}^2)$-branch for the second and a $(7.0523, 5.3293)$-branch for the first and third case which is sharp.

\subsubsection{Priority 8} 
 If we have chosen a variable $v$ with $\#_2(v)=4$ according to priority 8, such that $\#_2(N(v))<16$, then we have two distinct situations. By branching on $v$, we get at least a $(\{2\omega_4+2\omega_3+2\Delta_4\}^2)$-branch. (See Appendix~\ref{prooflem02}).           

\subsubsection{The 4- 5- and 6-regular case}
The part of the algorithm when we branch on variables of weight $h \neq 4$ will be called \emph{h-phase}. Branching according to priorities 4-8 is the \emph{4-phase}, according to priorities 9 and 10 the \emph{3-phase}. 
\begin{figure}
\centering
\psfrag{z}{$z$}
\psfrag{y}{$y$}
\psfrag{u1}{$u_1$}
\psfrag{u2}{$u_2$}
\psfrag{s1}{$s_1$}
\psfrag{s2}{$s_2$}
\psfrag{x}{$v$}
\psfrag{a}{$a$}
\psfrag{b}{$b$}
\psfrag{c}{$c$}
\psfrag{d}{$d$}
\psfrag{q}{$q$}
\psfrag{e}{$e$}
\psfrag{f}{$f$}
\psfrag{y}{$y$}
\psfrag{z}{$z$}
\subfigure[]{
    \label{prio62}
    \includegraphics[scale=0.595]{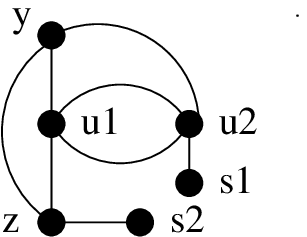}}
\subfigure[]{
    \label{prio63a}
    \includegraphics[scale=0.595]{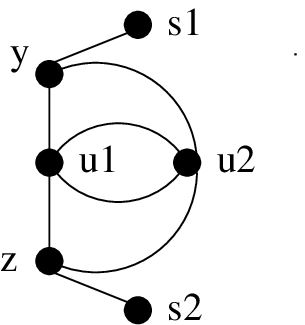}}
\subfigure[]{
    \label{prio63b}
    \includegraphics[scale=0.595]{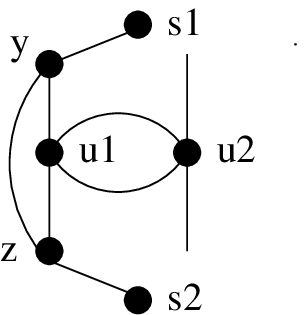}}
\subfigure[]{
    \label{prio63c}
    \includegraphics[scale=0.595]{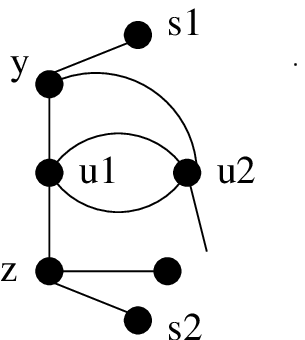}}
\subfigure[]{
    \label{wtr1}
    \includegraphics[scale=0.595]{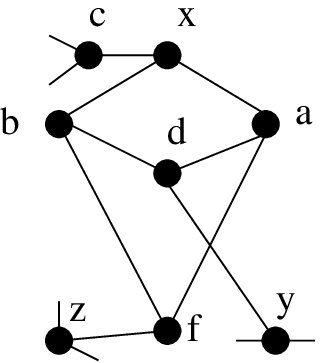}}
\subfigure[]{
    \label{abc1}
    \includegraphics[scale=0.595]{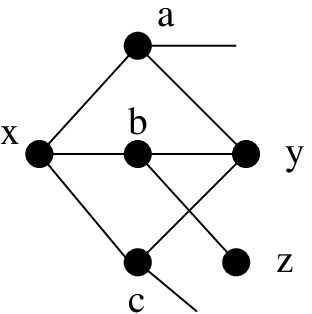}}
\caption{}
\label{fig1}
\end{figure}
In the following we have $4\le h \le 6$. Any $h$-regular  branch which was preceded by a branch from the $(h+1)$-phase can be neglected. This situation can only occur once on each path from the root to a leaf in the search tree. Hence, the run time is only affected by a constant multiple.
We now classify $h$-regular branches: An \emph{internal h-regular branch} is a $h$-regular branch such that another $h$-regular branch immediately follows in the search tree in at least one component. A \emph{final h-regular branch} is a $h$-regular branch such that no $h$-regular branch immediately succeeds in either of the components. When we are forced to do an $h$-regular branch, then according to {\bf HP} the whole graph must be $h$-regular at this point. 
\begin{obs}\label{regreduce}
If a branch is followed by a $h$-regular branch in one component, say in  $F[v]$, then in $F[v]$ any $u \in V(F)$ with $\#_2(u)<h$ will be reduced.  
\end{obs}
Due to Observation~\ref{regreduce} every vertex in $N(v)$ must be completely deleted  in $F[v]$. 
\begin{proposition}
\rut{1.1088^K} upper bounds any internal $h$-regular branch.
\end{proposition}
\begin{proof}
 By Observation~\ref{regreduce} for $h=4$ this yields at least a $(5\omega_4,\omega_4+4\Delta_4)$-branch as $v$ must have 4 different weight 4 neighbors due to {\bf HP}. If both components are followed by an $h$-regular branch we get a total reduction of $5\omega_4$ in both cases. 
The same way we can analyze internal $5$- and $6$-regular branches. This yields $(3\omega_5 ,\omega_5+5\Delta_5)$-,  $(\{3\omega_5\}^2)$-, $(3\omega_6,\omega_6+ 6\Delta_6)$- and $(\{3\omega_6\}^2)$-branches as for any $v \in V(F)$ we have $|N(v)|\ge 2$.
\qed
\end{proof} 
We now analyze a final $h$-regular $(\{b\}^2)$-branch with its preceding $(a_1,a_2)$-branch. The final $h$-regular branch might follow in the first, the second or both components of the $(a_1,a_2)$-branch. So, the \emph{combined analysis} would be a $(\{a_1+b\}^2,a_2)$, a $(a_1,\{a_2+b\}^2)$- and a $(\{a_1+b\}^2,\{a_2+b\}^2)$-branch.
\begin{proposition}
Any final $h$-regular branch ($h\in \{5,6\}$) considered together with its preceding branch can be upper bounded by $\Oh^*(1.1172^K)$.
\end{proposition}
\begin{proof}
We will apply a combined analysis for both branches.
Due to Observation~\ref{regreduce} $N(v)$  will be deleted in the corresponding component of the preceding branch. Due to Appendices~\ref{5reg} and \ref{6reg} the least amount we can get by deleting $N(v)$ is $\omega_5+\omega_4$ in case $h=5$ and $\omega_6+\omega_4$ in case $h=6$. Hence, we get four different branches: A $(\{3 \omega_5 + \omega_4 +5\Delta_5\}^2,\omega_5 + 5 \Delta)$-, a $(\{3 \omega_6 + \omega_4 +6\Delta_6\}^2,\omega_6 + 6 \Delta)$-, a  $(\{3 \omega_5 + \omega_4 +5\Delta_5\}^4)$- and a $(\{3 \omega_6 + \omega_4 +6\Delta_6\}^4)$-branch, respectively.
\qed
\end{proof}
\begin{proposition}
Any final $4$-regular branch considered with its preceding branch can be upper bounded by $\Oh^*(\rt)\approx$ \rut{1.11933^K}.
\end{proposition}
\begin{proof}
 We must analyze a final 4-regular branch together with any possible predecessor. These are all branches derived from priorities 4-8. See Appendix~\ref{OC4reg} for omitted cases.\\
{\bf Internal 4-regular branch} The two corresponding branches are a $(\{6 \omega_4 + 4 \Delta_4\}^2,\omega_4+4 \Delta_4)$-branch and a $(\{6 \omega_4 +4 \Delta_4\}^4)$-branch.\\
{\bf Priorities 4, 5 and 8} are all dominated by a $(\{2\omega_4+4\Delta_4\}^2)$-branch. Analyzing these cases together with a succeeding final 4-regular branch gives a $(\{3\omega_4+8\Delta_4\}^2,2\omega_4+4\Delta_4)$-branch and a  $(\{3\omega_4+8\Delta_4\}^4)$-branch.\\
{\bf Priority 7}
 Let $o$ be the number of weight 4 vertices from $N(v)$ and the 3- or 4-path, respectively. If in one component a final 4-regular branch follows then the worst case is when $o=0$ as any weight 4 vertex would be deleted completely and $\omega_4>\omega_3$. On the other hand if there is a component without an immediate 4-regular branch succeeding then the worst case appears when $o$ is maximal as $\omega_3 \ge \Delta_4$. So in the analysis we will consider for each case the particular worst case even though both together never appear.  \\
{\it 3-path with $u_0 \neq u_{l+1}$}:
First if there is a weight 4 variable in $N(u)$ we have at least the following branches: $a)$ $(\{3\omega_4+5\omega_3+4\Delta_4\}^2,\omega_4+\omega_3+3\Delta_4)$, $b)$ $(\omega_4+\omega_3+5\Delta_4,\{3\omega_4+3\omega_3+4\Delta_4\}^2)$ and $c)$ $(\{3\omega_4+5\omega_3+4\Delta_4\}^2,\{3\omega_4+3\omega_3+4\Delta_4\}^2)$. Any of those is upper bounded by $\Oh^*(\rt)$.
Now suppose for all $y \in N(v)$ we have $\#_2(y)=3$. Table~\ref{labb} captures the derived branches for certain combinations. Here we will also consider the weights of $u_0$ and $u_l$. 
\begin{table}
\centering
\begin{tabular}{|c|c|c|c|}\hline
$\#_2(u_0)$, $\#_2(u_{l+1})$    & left component & right component & both components\\ \hline
$\#_2(u_0)=3$&$(\{2\omega_4+6\omega_3+4\Delta_4\}^2,$ & $(\omega_4+6\omega_3,$ & $(\{2\omega_4+6\omega_3+4\Delta_4\}^2,$\\
 $\#_2(u_{l+1})=3$ & $\omega_4+4\omega_3)$    &$\{2\omega_4+6\omega_3+4\Delta_4\}^2)$ & $\{2\omega_4+6\omega_3+4\Delta_4\}^2)$ \\ \hline
$\#_2(u_0)=3$& $(\{3 \omega_4+5 \omega_3+4\Delta_4\}^2,$& $(\omega_4+5\omega_3+\Delta_4,$ & $(\{3 \omega_4+5 \omega_3+4\Delta_4\}^2,$\\
$\#_2(u_{l+1})=4$& $\omega_4+4\omega_3)$ &$ \{2\omega_4+5\omega_3+4\Delta_4\}^2)$& $ \{2\omega_4+5\omega_3+4\Delta_4\}^2)$\\ \hline 
$\#_2(u_0)=4$&$(\{4\omega_4+4\omega_3+4\Delta_4\}^2,$ & $(\omega_4+4\omega_3+2\Delta_4,$ &$(\{4\omega_4+4\omega_3+4\Delta_4\}^2,$\\
$\#_2(u_{l+1})=4$&$,\omega_4+4\omega_3)$& $\{2\omega_4+4\omega_3+4\Delta_4\}^2)$ &$\{2\omega_4+4\omega_3+4\Delta_4\}^2)$\\ \hline 
\end{tabular}
\caption{}
\label{labb}
\end{table}
Any entry is upper bounded by $\Oh^*(\rt)$ except $\alpha)$  $(\{2\omega_4+6\omega_3+4\Delta_4\}^2,\omega_4+4\omega_3)$  the left upper entry and $\beta)$ $(\omega_4+4\omega_3+2\Delta_4,\{2\omega_4+4\omega_3+4\Delta_4\}^2)$ the middle entry of the last row.
For $U \subseteq V(F)$ we define $E_3(U):=\{\{u,v\} \mid u \in U,\#_2(u)=3,v\not\in U\}$.
\begin{claim}
\begin{enumerate}
\item  Suppose for all $y \in Q:=N(v) \cup \{u_0,u_{l+1}\}$ we have $\#_2(y)=3$. Then there must be some $y' \in V \setminus (N(v) \cup \{u_0,u_{l+1}\})$ with $\#_2(y')=3$.
\item   Suppose for all $y \in N(v)$ we have $\#_2(y)=3$ and $\#_2(u_0)=\#_2(u_{l+1})=4$. Then there must be some $y' \in V \setminus (N(v) \cup \{u_0,u_{l+1}\})$ with $\#_2(y')=3$.
\end{enumerate}
\end{claim}
\begin{proof}
\begin{enumerate}
\item Assume the contrary.
For any $1 \le l \le 4$ we have $|E_3(Q \cup \{v\})|\le 10$. Due to \scp\ there is a weight 4 vertex $r$ adjacent to some vertex in $Q$. Observe that we must have $r \in Y$ as either there is $u \in N(v)$ with $u \in N(r)$ and $v \not \in N(r)$ or w.l.o.g. $u_0 \in N(r)$ but $u_1 \not \in N(r)$. Hence, $r$ has 4 weight 3 neighbors from $Q$ due to the choice of $v$. Hence we must have $|E_3(Q\cup \{v,r\})|\le 6$. Using the same arguments again we find some $r' \in Y$ with $|E_3(Q \cup \{v,r,r'\})|\le 2$. Again, due to \scp\ we find a $r'' \in Y$ with 4 weight 3 neighbors where at most two are from $Q$, a contradiction.
\item Assume the contrary.
Observe that $u_0,u_{l+1} \in Y$ and due to the choice of $v$ both have 4 weight 3 neighbors which must be from $N(v)$. From $|E_3(N[v])| \le 8$ follows that $|E_3(N[v]\cup \{u_0,u_{l+1}\})| =0$ which contradicts \scp. \qed 
\end{enumerate}
\end{proof}
Due to the last claim and Observation~\ref{regreduce} we have a $(\{2\omega_4+7\omega_3+4\Delta_4\}^2,\omega_4+4\omega_3)$-branch for case $\alpha)$ and 
a  $(\omega_4+4\omega_3+2\Delta_4,\{2\omega_4+5\omega_3+4\Delta_4\}^2)$-branch for case $\beta)$. Both are upper bounded by $\Oh^*(\rt)$.
\qed
\end{proof}


\subsection{The Cubic Case}
\subsubsection{Priority 9}
Observe that when we have arrived at this point, the graph $G_{var}$ must be 3-regular and each variable has three different neighbors, due to $G_{var}$ being reduced and due to Lemma \ref{lem0}.\ref{lem03a}.  Also, any 3-regular graph has an even number of vertices, because we have $3n=2m$.  Thus any branching must be of the form $(2i\cdot  \omega_3,2j \cdot \omega_3)$ for some $1 \le i,j$. Also, branching on any variable will at least result in a $(4\omega_3,4\omega_3)$-branch (see Lemma~\ref{lem0}.\ref{lem03a}). Note that any $u \in N(v)$ will be either set in $F[v]$ or in $F[\bar{v}]$, due to Lemma~\ref{lem0}.\ref{lem03b}.

\begin{lemma}\label{lem1}
Let $v$ have a pending triangle $a,b,c$ and $N(v)=\{a,p,q\}$.  Then by branching on $v$, we have an $(8 \omega_3,6 \omega_3)$-branch.
\end{lemma}

\begin{proof}
In $F[v]$ and $F[\bar{v}]$, the variables $a,b,c$ form a 3-quasi-lasso. Hence, due to Lemma~\ref{obs2}.\ref{lemlem} w.l.o.g., only $b$ remains in the reduced formula with $\#_2(b)=2$ (Lemma~\ref{obs2}~\ref{lemlem}). Also, in both branches, $q$ and $p$ are of weight two and therefore deleted. Note that  $N(\{q,p\})\cup  \{q,p\} \subseteq \{v,a,b,c,q,p\}$, contradicts \scp. Therefore, w.l.o.g., there is a variable $z \in N(q)$ such that $z \not \in \{v,a,b,c,q,p\}$.
So, in the branch where $q$ is set, also $z$ will be deleted. Thus, seven variables  will be deleted.\qed
\end{proof}

\subsubsection{Priority 10}
From now on, due to {\bf HP}, $G_{var}$ is  triangle-free and cubic. We show that if we are forced to choose a vertex $v$ to which none of the priorities 1-9 fits, we can choose $v$ such that we obtain either a   $(6\omega_3,8\omega_3)$- or a $(4\omega_3,10 \omega_3)$-branch.

\begin{lemma}\label{lem4}
Let $v$ be a vertex in $G_{var}$ and $N(v)=\{a,b,c\}$. Suppose that, w.l.o.g., in $F[v]$ $a,b$ and in $F[\bar{v}]$ $c$ will be set. Then we have a $(6\omega_3,8\omega_3)$-branch.
\end{lemma}
\begin{proof}
If $|(N(a) \cap N(b)) \setminus \{v\}| \le 1$, then by setting $a$ and $b$ in $F[v]$, five variables will be reduced. Together with $v$ and $c$, this is a total of  seven. If $|(N(a) \cap N(b)) \setminus \{v\}| =2$, then situation \ref{wtr1}  must occur (note the absence of triangles).  If $z=y$ then also $z\neq c$ due to \scp. Then in $F[v]$ due to Lemma~\ref{lem0}.\ref{lem01} $v,a,b,c,d,f,z$ will be deleted. If $z \neq y$ then $v,a,b,d,f,z,y$ will be deleted.
Together with $F[\bar{v}]$ where $c$ is set, we have a $(6\omega_3,8\omega_3)$-branch.
\qed
\end{proof}
\begin{lemma}
\label{lem5}
 If for any $v \in V(F)$ all its neighbors are set in one branch (say, in $F[v]$), we can perform  a $(6\omega_3,8\omega_3)$- or a $(4\omega_3,10 \omega_3)$-branch due to cubicity.
\end{lemma}
\begin{proof}
If $|N(a,b,c)\setminus \{v\}|\ge 5$, then in $F[v]$, 9 variables are deleted, so that we have a $(4\omega_3,10 \omega_3)$-branch. Otherwise, either one of the two following situations must occur:
$a)$ There is a variable $y\neq v$, such that $N(y)=\{a,b,c\}$, see Figure~\ref{abc1}. Then branch on $b$. In $F[\bar{b}]$, $v,y,a,c,z$ will disappear (due to {\bf RR-5} and Lemma~\ref{obs2}.\ref{lem02}). In $F[b]$, due to setting $z$, additionally a neighbor $f \not \in \{a, b,c,v,y\}$ of $z$ will be deleted due to \scp. This is a total of seven variables.\\
$b)$ There are variables $p,q$, such that $|N(p) \cap \{a,b,c\}|=|N(q)\cap \{a,b,c\}|=2$. The last part of Theorem 4.2 of \cite{KojKul2006} handles $b)$. See also Appendix~\ref{fullproof}.\qed
\end{proof}
Due to the last three lemmas, branchings according to priorities 9 and 10 are upper bounded by \rut{\rt}.
Especially, the $(4\omega_3,10 \omega_3)$-branch is sharp. 

\section{Combining Two Approaches}\label{comapp}
Kulikov and Kutzov~\cite{KulKut2007} achieved a run time of $\Oh^*(2^{\frac{K}{5.88}})$. This was obtained by speeding up the 5-phase by a concept  called 'clause learning'. As in our approach the 3- and 4-phase was improved we will show that if we use both strategies we can even beat our previous time bound. This means that in {\bf HP} we substitute priority 3 by their strategy with one exception: we prefer variables $v$ with a non weight 5 neighbor. Forced to violate this preference we do a simple branching of the form $F[v]$ and $F[\bar{v}]$.  For the analysis we redefine the measure $\gamma (F)$: 
we set $\omega_3= 0.9521$, $\omega_4=1.8320$, $\omega_5=2.488$ and keep the other weights. We call this measure $\tilde{\gamma} (F)$. We will reproduce the analysis of \cite{KulKut2007} briefly with respect to $\tilde{\gamma}(F)$ to show that their derived branches for the 5-phase are upper bounded by $\Oh^*(\rtt)$. It also can be checked that this is also true for the branches derived for the other phases by measuring them in terms of $\tilde{\gamma} (F)$, see Appendix~\ref{comapp246}.
Let $k_{ij}$ denote the number of weight $j$ variables occurring $i$ times in a 2-clause with some $v \in V(F)$ chosen for branching. Then we must have:
$k_{13}+k_{14}+k_{15}+2k_{24}+2k_{25}+3k_{35}=5$.
If $F'$ is the the formula obtained by assigning a value to $v$ and by applying the reduction rules afterwards we have:\\[2ex]
$\tilde{\gamma} (F) - \tilde{\gamma}(F')\ge 5\Delta_5 + \omega_5+ (\omega_3 - \Delta_5)k_{13}+(\Delta_4-\Delta_5)k_{14}+ (\frac{\omega_4}{2} - \Delta_5)2k_{24}
 +(\Delta_4-$\\[1ex] $\Delta_5)k_{25}  +(\frac{\omega_5}{2}-\frac{3}{2}\Delta_5)2 k_{35}
= 5.768+ 0.2961 k_{13} +  0.2239 (k_{14}+ k_{25}) +   0.26\cdot 2(k_{24}  +  k_{35})$\\[2ex]
Basically we reduce $\tilde{\gamma}(F)$ by at least $\omega_5+5\Delta_5$. Now the coefficients of the $k_{ij}$ in the above equation  express how  the reduction grows if $k_{ij}>0$.
If $k_{13}+k_{14}+2k_{24}+k_{25}+2k_{35} \ge 2$ we are done as $\tilde{\gamma} (F) - \tilde{\gamma}(F')\ge 6.2158$.\\
If $k_{13}=1$ and $k_{15}=4$ then \cite{KulKut2007} stated a $(5\Delta_5+\omega_5+(\omega_3-\Delta_5),5\Delta_5+\omega_5+(\omega_3-\Delta_5)+2 \Delta_5)$-branch and for $k_{25}=1$ and $k_{15}=3$ a $(5\Delta_5+\omega_5+(\Delta_4-\Delta_5),5\Delta_5+\omega_5+(\Delta_4-\Delta_5)+\omega_3)$-branch.
If $k_{14}=1$ and $k_{15}=4$ a branching of the kind $F[v],F[\bar{v},v_1],F[\bar{v},\bar{v}_1,v_2,v_3,v_4,v_5]$ is applied, where $\{v_1,\ldots,v_5\} = N(v)$.
From this follow a 
$(5\Delta_5+\omega_5+(\Delta_4-\Delta_5),4\Delta_5+\omega_5+\Delta_4+\omega_4+3\Delta_4+\Delta_5,5\omega_5+\omega_4)$- and a $(\omega_5+4\Delta_5+\Delta_4,\omega_5+4\Delta_5+\Delta_4+\omega_4+4\Delta_5,5\omega_5+\omega_4+3\omega_3)$-branch. This depends on whether $v_1$ has at least three neighbors of weight less than 5 in $F[\bar{v}]$ or not. We observed that we can get a additional reduction of $\Delta_5$ in the third component of the first branch as $N[v]$ cannot be a component in $V(F)$ after step 3 of Alg.~\ref{algo1} yielding a $(4\Delta_5+\omega_5+\Delta_4,5\Delta_5+\omega_5+4\Delta_4+\omega_4,5\omega_5+\omega_4 +\Delta_5)$-branch.
For the analysis of the 5-regular branch (i.e. $k_{15}=5$) we refer to Appendix~\ref{5regcomb}. It proceeds the same way as in the simple version of the algorithm except that we have to take into account the newly introduced branches. 

\begin{theorem}
{\sc Max-2-SAT} can be solved in time $\Oh^*(\rtt)\approx $ \rut{1.118^K}.
\end{theorem}

\section{Conclusion}
 We presented an algorithm solving {\sc Max-2-Sat} in $\Oh^*(\rtt )$, with $K$ the number of clauses of the input formula. This is currently the end of a sequence of polynomial-space algorithms  each improving on the run time: beginning with $\Oh^*(2^{\frac{K}{2.88}})$ which was achieved by \cite{NieRos2000a}, it was subsequently improved to \rut{2^{\frac{K}{3.742}}} by \cite{GraNie2000}, to \rut{2^{\frac{K}{5}}} by \cite{Graetal03a}, to \rut{2^{\frac{K}{5.217}}} by \cite{Kneetal2005}, to \rut{2^{\frac{K}{5.5}}} by \cite{KojKul2006} and finally to the hitherto fastest upper bound of \rut{2^{\frac{K}{5.88}}} by \cite{KulKut2007}. Our improvement has been achieved due to heuristic priorities concerning the  choice of variable for branching in case of a maximum degree four variable graph. As \cite{KulKut2007} improved the case where the variable graph has maximum degree five, it seems that the only way to speed up the generic branching algorithm is to improve the maximum degree six case.  Our analysis also implies that the situation when the variable graph is regular is not that harmful. The reason for this that the preceding branch must have reduced the problem size more than expected. Thus considered together these two branches balance each other. Though the analysis is to some extent sophisticated and quite detailed the algorithm has a clear structure. The implementation of the heuristic priorities for the weight 4 variables should be a straightforward task. Actually, we have already an implementation of Alg.~\ref{one}. It is still in an early phase but nevertheless the performance is promising. We are looking  forward to report on these results on another occasion.

\bibliographystyle{plain}

\newpage
\small
\begin{appendix}
\section{Additional Arguments Concerning 3-paths in the Non-regular Case}\label{Addarg}

\begin{proposition}
Let $v \in V(F)$ be chosen due to {\bf HP} such that $\#_2(v)=4$ and $v$ has a 3-path of length $l$ such that $u_0=u_{l+1}$. Then we have at least a $(\{\omega_4+3\omega_3+3\Delta_4\}^2)$-branch.
\end{proposition}
\begin{proof}
 In $F[v]$ and in $F[\bar{v}]$, $u_0u_1 \ldots u_l u_{l+1}$ is a lasso. So by Lemma~\ref{obs2}.\ref{lemlem}, $u_1,\ldots,u_l$ are deleted and the weight of $u_0$ drops by 2. If $\#_2(u_0)=4$ this yields a reduction of $l \cdot \omega_3+\omega_4$. If $\#_2(u_0)=3$ the reduction is $(l+1) \cdot \omega_3$ but then $u_0$ is set. 
If $N(u_0) \setminus N(v)$ is non-empty then we obtain a reduction of $\Delta_4$ in addition due to setting $u_0$. Otherwise there is a unique $r \in N(u_0) \setminus \{u_1,\ldots,u_l\}$ with $r \in N(v) \setminus \{u_1,\ldots,u_l\}$. If $\#_2(r)=4$  we get a $(\{2\omega_4+(l+1)\omega_3+(3-l)\Delta_4\}^2)$-branch. If $\#_2(r_1)=3$ then $r$ is set. As $(4-l)\le 2$ and by applying the same arguments to $r$ which previously where applied to $u_0$ we get at least a $(\{ \omega_4 + (l +1) \cdot \omega_3+(5-l)\Delta_4\}^2)$-branch. Observe that we used the fact that $\omega_4\ge 2\Delta_4$.\qed
\end{proof}

\section{Proof of the Statement in Priority 8 (Non-regular Case)}
\label{prooflem02}
\begin{figure}
\centering
\psfrag{x}{$v$}
\psfrag{a}{$a$}
\psfrag{b}{$b$}
\psfrag{c}{$c$}
\psfrag{d}{$d$}
\subfigure[]{
    \label{deg4a}
    \includegraphics[scale=0.6]{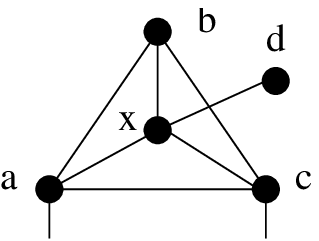}}
\subfigure[]{
    \label{deg4b}
    \includegraphics[scale=0.6]{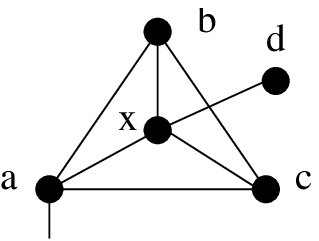}}
\subfigure[]{
    \label{deg4c}
    \includegraphics[scale=0.6]{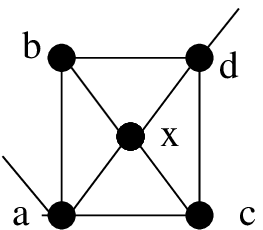}}
\subfigure[]{
    \label{deg4d}
    \includegraphics[scale=0.6]{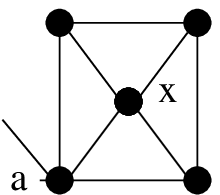}}

\caption{}
\label{figg1}
\end{figure}

\begin{proof}
Note that when we are forced to pick a variable $v$ according to priority 8, then either $v$ has four neighbors of weight 4 or for every weight 3 neighbor $z$ we have $N(z) \subseteq N(v)$. 
 From $\#_2(N(v))<16$ follows that, for every weight 3 neighbor $z$, we have $N(z) \subseteq N(v)$ due to the choice of $v$ according to {\bf HP}. Let $N^4$ (resp. $N^3$) be the set of weight four (three) neighbors of $v $. We analyze different cases induced by $k:=|N_3|$.\\
 Let $k=1$. If $N^3=\{b\}$, then there are vertices $a,c \in N^4$, such that $b \in N(a)$ and $b \in N(c)$. We must have $a \in N(c)$, or else $a$ would violate our assumption. Thus, we get the situation of Figure~\ref{deg4a}.\\
Let $k=2$. Then $N^3=\{b,c\}$ and assume that $b$ and $c$ are neighbors. If $b,c \in N(a)$ for $a \in N^4$, we have situation depicted in Figure~\ref{deg4b}. Otherwise $b \in N(a)$ and $c \in N(d)$ for $a,d \in N^4$. But then, priority 7 applies to both $a$ and $d$, which is a contradiction. In the case where $b$ and $c$ are not neighbors, it can be easily observed that we must have the situation in Figure~\ref{deg4c}, where  priority 7 applies to  $a$ and $d$.\\
 If $k=3$ it is easy to verify that we must have situation \ref{deg4d} in Figure \ref{fig1}. But then priority 7 applies to $a$.
 If $k=4$ then clearly $N[v]$ forms a component of five vertices which cannot appear after step 3 of Alg.~\ref{algo1}.\\
In Figure~\ref{deg4a} in either branch $F[v]$ or $F[\bar{v}]$, the variables $a,b,c$ form a 3-quasi-lasso, so by  Lemma~\ref{obs2}.\ref{lemlem} we get a reduction of $\omega_3+3 \omega_4 +\Delta_4 = 4 \omega_4$.\\ 
 In Figure~\ref{deg4b} in both branches the variables $a,b,c$ form a 3-lasso, so by Lemma~\ref{obs2}.\ref{lemlem} $b,c$ are deleted and $a$ is set due to  Lemma~\ref{lem0}.\ref{lem01}. We get a reduction of $ \omega_4+2 \omega_3$ from this. If $d\not \in N(a)$ we additionally get $2\Delta_4$, otherwise $\omega_4$. Altogether, we reduce $\gamma(F)$ by at least $2 \omega_4 +2 \omega_3 + 2\Delta_4$. \qed
\end{proof}

\section{Analysis of Internal 5- and 6-regular Branches}
\label{56reg}
We will consider branches which are immediately followed by a  $h$-regular branch in at least one component. In this component of the branch we can delete any variable in $N(v)$ additionally due to Observation~\ref{regreduce}. The Subsections~\ref{5reg} and~\ref{6reg} will explore  by how much  we additionally can decrement $\gamma(F)$ in the corresponding component in case $h \in \{5,6\}$. 
Let $k_{ij}$ denote the number of weight $j$ variables occurring $i$ times in a 2-clause with some $v \in V(F)$ chosen for branching.
\subsection{Internal 5-regular Branches}\label{5reg} 

\begin{proposition}
Let $v \in V(F)$ be the variable chosen due to {\bf HP} such that $\#_2(v)=5$. If this branch is followed by a $5$-regular branch in one component, then we can decrement $\gamma(F)$ by at least $\omega_5+\omega_4$ in addition to the weight of $v$ in that component.
\end{proposition}

\begin{proof}
  According to \cite{KulKut2007} we must have the following relation:
\begin{equation}\label{one}
k_{13}+k_{14}+k_{15}+2k_{24}+2k_{25}+3k_{35}=5
\end{equation}
We now have to determine an integer solution to (\ref{one}) such that $\omega_3 k_{13}+\omega_4 k_{14}+\omega_5 k_{15}+\omega_4 k_{24}+\omega_5 k_{25}+\omega_5 k_{35}$ is minimal. We can assume $k_{14}=k_{15}=0$ as we have $\omega_3< \omega_4 < \omega_5$. For any solution violating this property  we can find a smaller solution by setting $k'_{13}=k_{13}+k_{14}+k_{15}$, $k'_{14}=0$ and $k'_{15}=0$ and keeping the other coefficients. The same way we find that $k_{25}=0$ must be the case as $\omega_4 < \omega_5$.\\
If $k_{13}\ge 2$ we set $k'_{13}=k_{13} - 2\lfloor \frac{k_{13}}{2} \rfloor$, $k'_{24}=k_{24}+ \lfloor \frac{k_{13}}{2} \rfloor$ and keep the other coefficients.
By $2\omega_3 > \omega_4$ this is a smaller solution. Now suppose $k_{13}=1$, then we have $k_{24}=0$  in a minimal solution as $\omega_3 + \omega_4 > \omega_5$ (i.e., if $k_{24} \ge 1$ we set $k'_{13}=0$, $k'_{24}=k_{24}-1$ and $k'_{35}=k_{35}+1$) . But then no $k_{35}$ could satisfy (\ref{one}). Thus, we have $k_{13}=0$. Then the only solution is $k_{24}=1$ and $k_{35}=1$. Hence, the minimal reduction we get from $N(v)$ is $\omega_5+\omega_4$. \qed
\end{proof}

\subsection{Internal 6-regular Branches}\label{6reg}
\begin{proposition}
Let $v \in V(F)$ be the variable chosen due to {\bf HP} such that $\#_2(v)=6$. If this branch is followed by a $6$-regular branch in one component, then we can decrement $\gamma(F)$ by at least $\omega_6+\omega_4$ in addition to the weight of $v$ in that component.
\end{proposition}

\begin{proof}
In this case the following relation holds:
\begin{equation}\label{two}
k_{13}+k_{14}+k_{15}+k_{16}+2k_{24}+2k_{25}+2k_{26}+3k_{35}+3k_{36}+4k_{46}=6
\end{equation}
We now have to determine an integer solution to (\ref{two}) such that $\omega_3 k_{13}+\omega_4 k_{14}+\omega_5 k_{15}+\omega_6k_{16}+\omega_4 k_{24}+\omega_5 k_{25}+\omega_6 k_{26}+ \omega_5 k_{35} + \omega_6 k_{36} + \omega_6 k_{46}$ is minimal. As $\omega_3 < \omega_4 < \omega_5 < \omega_6$ we conclude that $k_{1\ell}=0$ for $4\le \ell  \le 6$, $k_{2\ell'}=0$ for $5\le \ell' \le 6$ and $k_{36}=0$. We also must have $k_{13} \le 1$ as in the section above. By $2\omega_4 > \omega_6$ we must have $k_{24} \le 1$. By (\ref{two}) we also have $k_{35}\le 2$ and $k_{46} \le 1$.\\
If $k_{13}=0$ the only solutions under the given restrictions are $k_{35}=2$ and $k_{24}=1,k_{46}=1$. If $k_{13}=1$ the only solution is $k_{35}=1,k_{24}=1$. Thus, the minimal amount we get by reduction from $N(v)$ is $\omega_6+\omega_4$. \qed
\end{proof}

\section{Omitted Cases of the Analysis of the Final 4-regular Case }
\label{OC4reg}
\begin{proposition}
Any final $4$-regular branch considered with its preceding branch can be upper bounded by $\Oh^*(\rt)$.
\end{proposition}
\begin{proof}
 Here we must analyze a final 4-regular branch together with any possible predecessor. These are all branches derived from priorities 4-8:
 Let $o$ be the number of weight 4 vertices from $N(v)$ or the 3- or 4-path, respectively. If in one component a final 4-regular branch follows then the worst case is when $o=0$ as any such vertex would be deleted completely and $\omega_4>\omega_3$. On the other hand if there is a component without an immediate 4-regular branch succeeding then the worst case appears when $o$ is maximal  as $\omega_3 \ge \Delta_4$. So in the analysis we will consider for each case the particular worst case even though both together never appear.  
\begin{description}
\item[Priority 6] Subcases $2$, $3(a)$ and $4$ of our non-regular priority-6 analysis can be analyzed similar to priorities 4, 5 and 8. We now analyze the remaining subcases.\\
{\it Subcase $1$} Here we deal with  small components which are directly solved without any branching. Therefore we get a $(\{3\omega_4+2\omega_3+4\Delta_4\}^2)$ -branch in the combined analysis. \\
Consider now cases $5$ and $3(b)$. Let $u_1,u_2$ be the picked limited pair. Due to {\bf HP} the variable $u_2$ has two weight 3 neighbors. Thus, if a final 4-regular branch is following in these cases we get a reduction of $2 \omega_3$ in addition (with respect to the component of the branch). For both cases we derived a non symmetric branch, e.g., an $(a,b)$-branch with $a \neq b$. Depending whether the final 4-regular branch follows in the first, the second or both components we derive three combined branches: $a)$ $(\{3\omega_4+4\omega_3+4 \Delta_4\}^2,2\omega_3+\omega_4+2\Delta_4)$, $b)$ $(2\omega_3+2\omega_4+2\Delta_4,\{3\omega_4+4\omega_3+4\Delta_4\}^2)$- and $c)$ $(\{3\omega_4+4\omega_3+4\Delta_4\}^2,\{3\omega_4+4\omega_3+4\Delta_4\}^2)$. As $\Oh^*(\rt)$ not proper upperbounds $a)$ we need a further discussion for the two subcases. Remember that in the first component of $a)$ some weight 3 neighbor $t$ of $v$ is set.\\
{\it Subcase 3b} First suppose that $N(z)\setminus (N(u_1)\cup N(u_2))=\emptyset$ and  $N(y)\setminus (N(u_1)\cup N(u_2))=\emptyset$, see Figure~\ref{waga}. Then by either branching on $y$ or $z$ we get a $(\{2\omega_4+4\omega_3\}^2)$-branch. In this case the combined analysis is similar to priorities 4, 5 and 8. Secondly, w.l.o.g. we have $N(z)\setminus (N(u_1)\cup N(u_2))\neq\emptyset$, see Figure~\ref{waga2} and \ref{waga3}. In Figure~\ref{waga2} we might have picked $y=v$ or $z=v$. But observe that in both cases in the branch where the particular weight 3 neighbor $t$ is set ($t=s$ if $v=z$ and $t=z$ if $v=y$) such that in this component a 4-regular branch follows we have a $(\{3\omega_4+5\omega_3+4\Delta_4\}^2,\omega_4+2\omega_3+2\Delta_4)$-branch in the combined analysis instead of $a)$. If the case in Figure~\ref{waga3} matches then we have $t=s$. Then in the branch were $s$ is set $y$ and $u_1$ will be reduced due to {\bf RR-5} and $N[u_2]\setminus \{u_1\}$ due to the fact that a 4-regular branch follows. Thus, the derived branch is the same as for the case of Figure~\ref{waga2}.\\
{\it Subcase 5} As the vertices in $N(u_1)\cup N(u_2)$ can not form a component w.l.o.g. we have that $N(z)\setminus (N(u_1)\cup N(u_2))\neq \emptyset$. In this case we branch on $u_1$. Now in the branch where we set $z$ (i.e., $z=t$) such that a 4-regular branch follows in that component we have a $(\{3\omega_4+5\omega_3+4\Delta_5\}^2,2\omega_4+2\omega_3)$-branch in the combined analysis insted of $a)$\\
Both branches replacing $a)$ have an upper bound of $\Oh^*(\rt)$.\\
\begin{figure}
\centering
\psfrag{z}{$z$}
\psfrag{y}{$y$}
\psfrag{u1}{$u_1$}
\psfrag{u2}{$u_2$}
\psfrag{s1}{$s_1$}
\psfrag{s2}{$s_2$}
\psfrag{x}{$v$}
\psfrag{a}{$a$}
\psfrag{b}{$b$}
\psfrag{c}{$c$}
\psfrag{d}{$d$}
\psfrag{s}{$s$}
\subfigure[]{
    \label{waga}
    \includegraphics[scale=0.595]{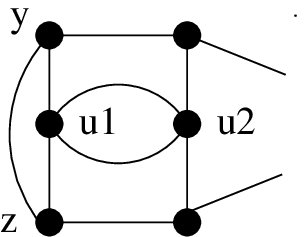}}
\subfigure[]{
    \label{waga2}
    \includegraphics[scale=0.595]{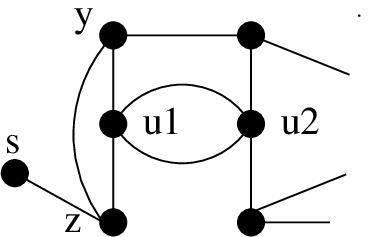}}
\subfigure[]{
    \label{waga3}
    \includegraphics[scale=0.595]{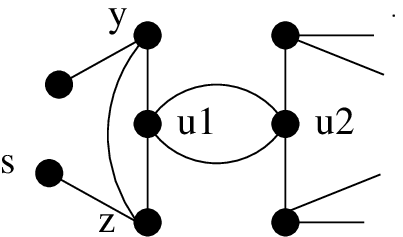}}
\caption{}
\end{figure}

\item[Priority 7] 
{\it 4-path} In this case we have the following branches: $a)$ $(\{3\omega_4+4\omega_3+4\Delta_4\}^2,\omega_4+\omega_3+3\Delta_4)$, $b)$ $(2\omega_4+\omega_3+3\Delta_4,\{3\omega_4+3\omega_3+4\Delta_4\}^2)$, $c)$ $(\{3\omega_4+4\omega_3+4\Delta_4\}^2,\{3\omega_4+3\omega_3+4\Delta_4\}^2)$. The cases $a)$ and $c)$ are not upper bounded by $\Oh^*(\rt)$ and hence need further discussion.\\
 Suppose there is a vertex $y \in D :=N(v) \cup \{u_0,\ldots,u_{l-1}\}$ with weight 4. Then by Observation~\ref{regreduce} we have branches $a')$ $(\{4\omega_4+3\omega_3+4\Delta_4\}^2,\omega_4+\omega_3+3\Delta_4)$ and $c')$ $(\{4\omega_4+3\omega_3+4\Delta_4\}^2,\{3\omega_4+3\omega_3+4\Delta_4\}^2)$ which are both upperbounded by $\Oh^*(\rt)$. For the remaining case we need the next proposition.
For $U \subseteq V(F)$ we define $E_3(U)=\{\{u,v\} \mid u \in U,\#_2(u)=3,v\not\in U\}$. 
\begin{claim}\label{no4}
Suppose for all $y \in D$ we have $\#_2(y)=3$. Then there must be some $y' \in V \setminus (D \cup \{v,u_l\})$ with $\#_2(y')=3$.
\end{claim}

\begin{proof} Assume the contrary. Observe that if $l \ge 3$ then $u_l \in Y$ due to $u_{l-2} \not \in N(u_l)$. If $l=2$ and $u_0 \not \in  N(u_2)$ then also $u_l \in Y$ holds. Let us assume this case as the other one will be treated separately.\\
Now due to the choice of $v$  we have that $u_l$ must be adjacent to $v,u_{l-1}$ and to two further weight 3 vertices in $D$. Therefore  and as $D \cup\{v,u_l\}$   can not be a component we have $l<4$. 
 Also for any $2 \le l \le 3$ we always have $|E_3(D \cup\{v,u_l\})|\le 8-2l$. There must some weight 4 vertex $r \not \in D\cup\{v, u_l\}$ adjacent to some weight 3 vertex in $D$ as we have no small components. Note that $r \in Y$, as $v \not \in N(r)$ or w.l.o.g. $u_0 \in N(r)$ but $u_1 \not \in N(r)$. Due to the choice of $v$, $r$ must have at least three weight 3 neighbors. Hence $l=2$.
If $r$ has 4 weight 3 neighbors then $(D \cup\{v,u_l,r\})$ forms a component which is a contradiction.
Hence, we have $|E_3(D \cup\{v,u_l,r\})|= 1$ and therefore we find again some $r' \in Y \setminus (D \cup \{v,u_l,r\})$  which is adjacent to at least 3 weight 3 vertices where at most one is from $D$. Thus there must be some weight 3 vertex in   $V \setminus (D\cup \{v,u_l\})$, a contradiction.\\
Now suppose $l=2$ and $u_0 \in N(u_2)$. Let $N(u_0)=\{z,u_1,u_2\}$. If $z \not \in N(u_2)$ then $u_2 \in Y$ and the first part of the proof applies.
Now suppose $z \in N(u_2)$ and $\#_2(z)=3$. Then it follows that $z \in N(v)$ and $|E_3(\{D\cup \{v,u_2\}\})|\le 2$. Now due to \scp\ we can find an $r \in Y \setminus (D \cup \{v,u_l\})$ which is adjacent to at least three weight 3 vertices where only two can be from $D\cup \{v,u_l\}$, a contradiction.\\
Suppose $z \in N(u_2)$ and $\#_2(z)=4$. As $u_1 \not \in N(z)$ we have $z \in Y$. Thus, $z$ is adjacent to the two weight 3 vertices in $N(v) \setminus \{u_2,u_1\}$. As $D \cup\{v,u_2,z\}$ is not a component we have $|E_3(\{D \cup\{v,u_2,z\}\})|=2$. Similarly, a contradiction follows.
\qed
\end{proof}
If for all $y \in D$ we have $\#_2(y)=3$ from the last claim and Lemma~\ref{obs2}.\ref{lemlem} we can derive two branches $a'')$ $(\{3\omega_4+5\omega_3+4\Delta_4\}^2,\omega_4+\omega_3+3\Delta_4)$ and $c'')$ $(\{3\omega_4+5\omega_3+4\Delta_4\}^2,\{3\omega_4+3\omega_3+4\Delta_4\}^2)$ which are upper bounded by $\Oh^*(\rt)$.

{\it 3-path} In the case of 3-path such that $u_0=u_{l+1}$ the branch with $l=2$ is dominated by all other choices. Since this is a $(\{7.21\}^2)$-branch we refer to priorities 4, 5 and 8 from above.\\
\end{description}
\qed
\end{proof}

\section{Full Proof of Lemma~\ref{lem5}}\label{fullproof}

\begin{figure}
\psfrag{x}{$v$}
\psfrag{a}{$a$}
\psfrag{b}{$b$}
\psfrag{c}{$c$}
\psfrag{d}{$d$}
\psfrag{q}{$q$}
\psfrag{e}{$e$}
\psfrag{f}{$f$}
\psfrag{y}{$y$}
\psfrag{z}{$z$}
\psfrag{x}{$v$}
\psfrag{a}{$a$}
\psfrag{b}{$b$}
\psfrag{c}{$c$}
\psfrag{d}{$d$}
\psfrag{y}{$y$}
\psfrag{p}{$p$}
\psfrag{q}{$q$}
\psfrag{z}{$z$}
\psfrag{h1}{$h_1$}
\psfrag{h2}{$h_2$}
\psfrag{h3}{$h_3$}
\psfrag{h4}{$h_4$}
\centering
\subfigure[]{
    \label{abc2}
    \includegraphics[scale=0.595]{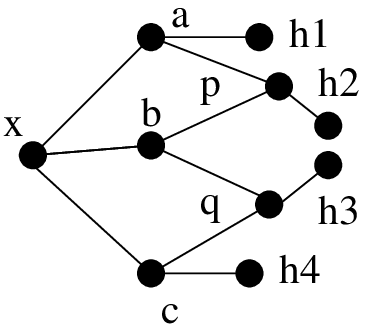} }
\subfigure[]{
    \label{abc3}
    \includegraphics[scale=0.595]{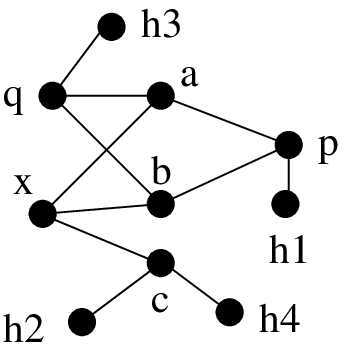} }
\caption{}
\label{fig2}
\label{fig3}
\end{figure}

\begin{proof}
If $|N(a,b,c)\setminus \{v\}|\ge 5$, then in $F[v]$, 9 variables are deleted, so that we have a $(4\omega_3,10 \omega_3)$-branch. Otherwise, either one of the two following situations must occur:
$a)$ There is a variable $y\neq v$, such that $N(y)=\{a,b,c\}$, see Figure~\ref{abc1}. Then branch on $b$. In $F[\bar{b}]$, $v,y,a,c,z$ will disappear (due to {\bf RR-5} and Lemma~\ref{obs2}.\ref{lem02}). In $F[b]$, due to setting $z$, additionally a neighbor $f \not \in \{a, b,c,v,y\}$ of $z$ will be deleted as the vertices $a,b,c,v,y,z$ do not form a component. This is a total of seven variables.\\
$b)$ There are variables $p,q$, such that $|N(p) \cap \{a,b,c\}|=|N(q)\cap \{a,b,c\}|=2$, see Figure \ref{abc2} and \ref{abc3}. In $F[v]$, the variables $a,b,c,p,q$ will be set. Then, at least 3 additional variables will be deleted (even if there are $1 \le i<j\le 4$ with $h_i=h_j$). Theorem 4.2 of \cite{KojKul2006} contains also an alternative proof of $b)$.
\qed
\end{proof}

\section{Additional Analysis of the Combined Approach}
\subsection{5-regular Branches in the Combined Approach}\label{5regcomb}
Internal 5-regular branches yield the same recurrences as in the simple approach. Final 5-regular branches must be analyzed together with their immediate preceding branch. Thus they have to be analyzed together with the introduced branches of \cite{KulKut2007}. Table~\ref{tab3} captures some cases ($k_{15}=5$; $k_{13}=1, k_{14}=4$; $k_{25}=1, k_{15}=3$).
\begin{table}
\centering
\begin{tabular}{|c|c|c|c|}\hline
  case    & one component & both components  & upper bound\\ \hline 
$k_{15}=5$& $(\{7\omega_5+5\Delta_5\}^2,\omega_5+5\Delta_5)$&$(\{7\omega_5+5\Delta_5\}^4)$&$\Oh^*(1.0846^K)$  \\ \hline 
$k_{13}=1$, $k_{15}=4$& $(\{6\omega_5+\omega_3 +5\Delta_5\}^2,\omega_5+4\Delta_5+\omega_3)$ & $\{6\omega_5+\omega_3 +5\Delta_5\}^4)$&\rut{1.0878^K} \\ \hline
$k_{25}=1$, $k_{15}=3$& $(\{6\omega_5+5\Delta_5\}^2,\omega_5+3\Delta_5+(\omega_5-\omega_3)$ &$(\{6\omega_5+5\Delta_5\}^4)$ & \rut{1.0914^K} \\ \hline
\end{tabular}
\caption{}
\label{tab3}
\end{table}
For the case $k_{14}=1$ and $k_{15}=4$ there are two recurrences for the branching $F[v],F[\bar{v},v_1],F[\bar{v},\bar{v}_1,v_2,v_3,v_4,v_5]$. The first recurrence assumes that $v_1$ has at least three neighbor of weight less than five in $F[\bar{v}]$: {\bf (A)} $(5\Delta_5+\omega_5+(\Delta_4-\Delta_5),4\Delta_5+\omega_5+\Delta_4+\omega_4+3\Delta_4+\Delta_5,5\omega_5+\omega_4+\Delta_5)$. The other {\bf (B)} $(\omega_5+4\Delta_5+\Delta_4,\omega_5+4\Delta_5+\Delta_4+\omega_4+4\Delta_5,5\omega_5+\omega_4+3\omega_3)$ captures the remaining case.  Both branches have three components. Table~\ref{tab4} captures the combined branches of a immediately following final 5-branch and branches {\bf (A)} and {\bf (B)}. This depends on whether the final 5-regular branch follows after the first $(1)$, the second $(2)$ or the third $(3)$ component or in any combination of them.\\
\begin{table}[H]
\begin{tabular}{|c|c|c|c|}\hline 
Branch-type &components & combined branch & upper bound\\ \hline 
{\bf (A)} & $(1)$&$(\{6\omega_5+\omega_4+5\Delta_5\}^2,$&\rut{1.0912^K}\\ 
& & $5\Delta_5+\omega_5+(\Delta_4-\Delta_5)+\omega_4+3\Delta_4+\Delta_5,$&\\
& & $5\Delta_5+\omega_5+(\Delta_4-\Delta_5)+4\omega_4+\omega_3+\Delta_5)$ &\\ \hline 
{\bf (A)}& $(2)$ & $(5\Delta_5+\omega_5+(\Delta_4-\Delta_5),\{6\omega_5+\omega_4+5\Delta_5\}^2,$ &\rut{1.1094^k}\\ 
& & $5\Delta_5+\omega_5+(\Delta_4-\Delta_5)+4\omega_4+\omega_3+\Delta_5)$& \\ \hline
{\bf (A)} &$(3)$ & $(5\Delta_5+\omega_5+(\Delta_4-\Delta_5),$ &\rut{1.1175^K}\\ 
& & $5\Delta_5+\omega_5+(\Delta_4-\Delta_5)+\omega_4+3\Delta_4+\Delta_5,$& \\ 
& & $\{6\omega_5+\omega_4+5\Delta_5+\omega_3\}^2)$ &\\ \hline 
{\bf (B)} & $(1)$ & $(\{6\omega_5+\omega_4+5\Delta_5\}^2,$& \rut{1.0894^K}\\
& & $5\Delta_5+\omega_5+(\Delta_4-\Delta_5)+\omega_4+4\Delta_5,$&\\
& &$5\Delta_5+\omega_5+(\Delta_4-\Delta_5)+4\omega_4+\omega_3+3\omega_3)$& \\ \hline
{\bf (B)} &$(2)$ &$(5\Delta_5+\omega_5+(\Delta_4-\Delta_5),\{6\omega_5+\omega_4+5\Delta_5\}^2,$&\rut{1.1052^K}\\
& &$5\Delta_5+\omega_5+(\Delta_4-\Delta_5)+4\omega_4+\omega_3+3\omega_3)$&\\ \hline
{\bf (B)} &$(3)$ & $(5\Delta_5+\omega_5+(\Delta_4-\Delta_5),$&\rut{1.1159^K}\\ 
& & $5\Delta_5+\omega_5+(\Delta_4-\Delta_5)+\omega_4+4\Delta_5,$&\\
& &$\{6\omega_5+\omega_4+5\Delta_5+3\omega_3\}^2) $&\\ \hline
{\bf (A)} or& $(1)(2)$/$(1)(3)$& $(\{6\omega_5+\omega_4 +5\Delta_5\}^4,\omega_5+4\Delta_5+\Delta_4)$&\rut{1.1126^K}\\ 
 {\bf (B)} &/$(2)(3)$ & &  \\ \hline 
{\bf (A)} or & $(1)(2)(3)$ &$ (\{6\omega_5+\omega_4 +5\Delta_5\}^6)$&\rut{1.0936^K} \\ 
{\bf (B)} & & & \\ \hline
\end{tabular}
\caption{The second column indicates after which components a final 5-regular branch immediately follows.}
\label{tab4}
\end{table}\noindent
We would like to comment the recurrences in the third and sixth row. Here we get a reduction of $\omega_3$ and $3\omega_3$ in addition to $5\omega_4+\omega_4$ from $v,v_1,\ldots, v_5$. This additional amount comes from clauses $C$ such that $|C \cap \{v,v_1\ldots,v_5\}|=1$. Especially in the first case due to \scp\ $N[v]$ is not  component and thus at least one further variable must be deleted. 
\begin{proposition}\label{restprop}
Let $v \in V(F)$ be the variable chosen for branching by Alg.~\ref{algo1} such that $\#_2(v)=5$. Assume $v$ induces a solution to  equation~(\ref{one}) such that it is different from  $k_{13}=1,k_{15}=4$; $k_{15}=5$; $k_{25}=1,k_{15}=3$; $k_{14}=1,k_{15}=4$ ($\star$). If a $5$-regular branch follows in one component we have at least a $(\{3\omega_5+\omega_4+5\Delta_5\}^2,\omega_5+3\Delta_5+2\Delta_4)$-branch and if it follows in both a $(\{3\omega_5+\omega_4+5\Delta_5\}^4)$-branch.
\end{proposition}
\begin{proof}
If a component is followed by a final 5-regular branch the least amount we get by reduction from $N(v)$ is $\omega_5+\omega_4$. This refers to the case $k_{35}=1$ and $k_{24}=1$ which follows analogously  from Section~\ref{5reg}. \\
The least reduction from $N(v)$ without a following final 5-regular branch can be found as follows: Consider any solution of equation~(\ref{one}) expect the ones in ($\star$). Among them find  one which minimizes
\begin{equation}\label{three}
\Delta_3k_{13}+\Delta_4k_{14}+\Delta_5k_{15}+(\Delta_4+\Delta_3)k_{24}+(\Delta_5+\Delta_4)k_{25}+(\Delta_5+\Delta_4+\Delta_3)k_{35}
\end{equation}
We can assume $k_{24}=0$ as we have $\Delta_4+\Delta_3>\Delta_5+\Delta_4$. 
As we are excluding ($\star$) we must have $k_{15} \le4$. If $k_{15}=4$ we conclude that either $k_{14}=1$ oder $k_{13}=1$. Both solutions are forbidden (see ($\star$)). Thus we must have $k_{15} \le 3$.\\ 
If $k_{35}=1$ then there is a better solution as $\Delta_5+2\Delta_4<\Delta_5+\Delta_4+\Delta_3$: set $k'_{35}=0,k'_{15}=k_{15}+1,k'_{14}=k_{14}+2$ and keep the other coefficients. Note that in this case we must have $k_{15}\le 2$ which assures that the new solutions is different from the ones in ($\star$). Therefore it follows that $k_{35}=0$.\\
Now suppose $k_{25}=2$, then $k_{15}=1$ holds. But then $k'_{25}=k_{25}-1,k'_{15}=k_{15}+1,k'_{14}=1$ is a no worse solution. Thus $k_{25} \le 1$. If $k_{25}=1$ then with $k_{15}=2$ and $k_{14}=1$ this is minimal (since $k_{15}=3$ is forbidden ($\star$)).\\
Suppose $k_{25}=0$, then clearly the best solution is $k_{15}=3$ and $k_{14}=2$. Both solutions provide a reduction of $3 \Delta_5+ 2 \Delta_4$ which is minimal. \\
Now we analyze a final 5-regular branch and a branch different form ($\star$) satisfying equation~(\ref{one}). If the final 5-regular branch follows in only one component then we have at least a $(\{3\omega_5+\omega_4+5\Delta_5\}^2,\omega_5+3\Delta_5+2\Delta_4)$-branch in the combined analysis. If it follows in both then a $(\{3\omega_5+\omega_4+5\Delta_5\}^4)$-branch upper bounds correctly.
\qed
\end{proof}
Due to Proposition~\ref{restprop} we can upper bound the 5-regular branches  whose predecessors are different from  $k_{13}=1,k_{15}=4$; $k_{15}=5$; $k_{25}=1,k_{15}=3$; $k_{14}=1,k_{15}=4$ by \rut{1.1171^K} in their combined analysis.

\subsection{Analysis  of the 6- 4- and 3-phase in the Combined Approach}\label{comapp246}
Here we provide the run times under $\tilde{\gamma}(F)$ for the cases we did not consider in Section~\ref{comapp}. The run time has been estimated with respect to $\tilde{\gamma}(F)$. Names will refer to the corresponding ones in the analysis of Alg.~\ref{algo1}.

\subsubsection{Non-regular Branches} In Table~\ref{tabb4} we find the derived recurrences for each priority of {\bf HP} if we have a non-regular branch. You can find them together with their run times. Priority 3 is not considered as the 5-phase has been analyzed in Sections~\ref{comapp} and~\ref{5regcomb}.
\begin{table}\centering
\begin{tabular}{|c|c|c|}\hline
Priorities& branch &upper bound \\ \hline
{\bf Priority 1}& $(7,7)$& \rut{1.1042^K}\\ \hline
{\bf Priority 2}&$(\{\omega_6+5\Delta_6+\Delta_5\}^2)$ & \rut{1.118^K}\\ \hline
{\bf Priority   4}&$(\{2\omega_4+4\Delta_4\}^2)$&\rut{1.102^K}\\ \hline
{\bf Priority 5}&$(\{3\omega_4+2\omega_3\}^2)$&\rut{1.1099^K} \\ \hline
{\bf Priority 6} &$(2\omega_4+2\omega_3+2\Delta_4,\omega_4+2\omega_3+2\Delta_4)$&\rut{1.1143^K}\\ \hline
{\bf Priority 7}& $(\omega_4+\omega_3+5\Delta_4,\omega_4+\omega_3+3\Delta_4)$&\rut{1.1172^K}\\ 
                       &$(\{\omega_4+3\omega_3+3\Delta_4\}^2)$&\rut{1.1^K}\\ 
                      &$(2\omega_4+\omega_3+3\Delta_4,\omega_4+\omega_3+3\Delta_4)$&\rut{1.1165^K}\\ \hline 
{\bf Priority 8}&$(\{2\omega_4+2\omega_3+2\Delta_4\}^2)$&\rut{1.1^K} \\ \hline
{\bf Priority 9}&$(8\omega_3,6\omega_3)$&\rut{1.1105^K}\\ \hline
{\bf Priority 10}&$(8\omega_3,6\omega_3)$&\rut{1.1105^K}\\
                         &$(4\omega_3,10\omega_3)$&\rut{1.118^K}\\ \hline
\end{tabular}
\caption{The non-regular cases}
\label{tabb4}
\end{table}

\subsubsection{Regular Branches}
Table~\ref{tabb5} captures the run times of any internal 6, 5 or 4-regular branch. Table~\ref{tabb6} considers final 4 or 6-regular branches together with their preceding branches.  The case where we have chosen $v$ due to priority 7 such that $v$ has a 3-path with $u_0\neq u_l$ is treated separately.
\begin{table}
\centering
\begin{tabular}{|c|c|c|}\hline
case & branch & upper bound \\ \hline
{\bf Internal 6-regular}&$(3\omega_6,\omega_6+6\Delta_6)$&\rut{1.0978^K}\\
& $(\{3\omega_6\}^2)$&\rut{1.0802^K}\\ \hline
{\bf Internal 5-regular}&$(3\omega_5,\omega_5+5\Delta_5)$&\rut{1.1112^K}\\
& $(\{3\omega_5\}^2)$&\rut{1.0974^K}\\ \hline
{\bf Internal 4-regular}&$(5\omega_4,\omega_4+4\Delta_4)$&\rut{1.103^K}\\
& $(\{5\omega_4\}^2)$&\rut{1.079^K}\\ \hline
\end{tabular}
\caption{Internal $h$-regular cases ($h \in \{4,5,6\}$) an their upper bounds.}
\label{tabb5}
\end{table}

\begin{table}
 \centering
\begin{tabular}{|c|c|c|}\hline
Preceding branch & branch & upper bound \\ \hline
\multicolumn{3}{|c|}{{\bf Final 6-regular Branch}}\\ \hline
Any 6-phase branch &$(\{3\omega_6+\omega_4+6\Delta_6\}^2,\omega_6+6\Delta_6)$&\rut{1.11^K}\\
&$(\{3\omega_6+\omega_4+6\Delta_6\}^4)$&\rut{1.105^K}\\ \hline
\multicolumn{3}{|c|}{{\bf Final 4-regular Branch}}\\ \hline
{\bf Internal 4-regular}& $(\{6\omega_4+4\Delta_4\}^2,\omega_4+4\Delta_4)$&\rut{1.1115^K}\\
& $(\{6\omega_4+4\Delta_4\}^4)$&\rut{1.1003^K}\\ \hline
{\bf Priorities 4,5 and 8}&$(\{3\omega_4+8\Delta_4\}^2,2\omega_4+4\Delta_4)$&\rut{1.1115^K}\\
Cases $2$,$3(a)$,$4$ of {\bf Priority 6} &$(\{3\omega_4+8\Delta_4\}^4)$ &\rut{1.117^K} \\ \hline
{\bf Priority 6} & &\\
Case 1&$(\{3\omega_4+2\omega_3+4\Delta_4\}^2)$&\rut{1.07^K}\\
Case $5$,$3(b)$ &$(\{2\omega_4+2\omega_3+2\Delta_4,\{3\omega_4+4\omega_3+4\Delta_4\}^2)$\ \ $b)$&\rut{1.109^K}\\
$b)$ and $c)$ of the analysis&$(\{3\omega_4+4\omega_3+4\Delta_4\}^4)$\ \ $c)$&\rut{1.1143^K}\\
Case $3b)$, case $a)$ of the analysis&$(\{3\omega_4+4\omega_3+5\Delta_4\}^2, \omega_4+2\omega_3+2\Delta_4)$&\rut{1.1151^K}\\ 
Case $5$, case $a)$ of the analysis&$(\{3\omega_4+4\omega_3+5\Delta_4\}^2, 2\omega_4+2\omega_3)$&\rut{1.1145^K}\\ \hline 
{\bf Priority 7}& &  \\
Case of a 4-path& & \\
Case $b)$ & $(2\omega_4+\omega_3+3\Delta_4,\{3\omega_4+3\omega_3+4\Delta_4\}^2)$&\rut{1.1155^K}\\
Case $a')$ & $(\{4\omega_4+3\omega_3+4\Delta_4\}^2,\omega_4+\omega_3+3\Delta_4)$&\rut{1.1156^K}\\
Case $c')$ & $\{4\omega_4+3\omega_3+4\Delta_4\}^2,\{3\omega_4+3\omega_3+4\Delta_4\}^2)$&\rut{1.115^K}\\
Case $a'')$ & $(\{3\omega_4+5\omega_3+4\Delta_4\}^2,\omega_4+\omega_3+3\Delta_4)$&\rut{1.1152^K}\\
Case $c'')$ &$(\{3\omega_4+5\omega_3+4\Delta_4\}^2,(\{3\omega_4+3\omega_3+4\Delta_4\}^2)$&\rut{1.1147^K}\\ 
Case of a 3-path with $u_0=u_{l+1}$ & similar to {\bf priorities 4,5 and 8} & \\ \hline
\end{tabular}
\caption{The final $h$-regular cases ($h \in \{4,6\}$) and their combined analysis}
\label{tabb6}
\end{table}
\paragraph{3-path}
Finally we consider the case when a variable chosen to priority 7 has a 3-path with $u_0\neq u_{l+1}$. The cases $a)$ $(\{3\omega_4+5\omega_3+4\Delta_4\}^2,\omega_4+\omega_3+3\Delta_4)$, $b)$ $(\omega_4+\omega_3+5\Delta_4,\{3\omega_4+3\omega_3+4\Delta_4\}^2)$ and $c)$ $(\{3\omega_4+5\omega_3+4\Delta_4\}^2,\{3\omega_4+3\omega_3+4\Delta_4\}^2)$ are upper bounded by \rut{1.1152^K}, \rut{1.1159^K} and \rut{1.1147^K}.
Table~\ref{tabb1} captures the branches together with their run times in the combined algorithm if for all $y \in N(v)$ we have $\#_2(y)=3$.
\begin{table}
\centering
\begin{tabular}{|c|c|c|c|}\hline
 $\#_2(u_0)$, $\#_2(u_{l+1})$         & left component & right component & both components\\ \hline
$\#_2(u_0)=3$& case $\alpha$ instead  & $(\omega_4+6\omega_3,$ & $(\{2\omega_4+6\omega_3+4\Delta_4\}^2,$\\
 $\#_2(u_{l+1})=3$ &   &$\{2\omega_4+6\omega_3+4\Delta_4\}^2)$ & $\{2\omega_4+6\omega_3+4\Delta_4\}^2)$ \\ 
 upper bounds & & \rut{1.1075^K}&\rut{1.1136^K} \\ \hline
$\#_2(u_0)=3$& $(\{3 \omega_4+5 \omega_3+4\Delta_4\}^2,$& $(\omega_4+5\omega_3+\Delta_4,$ & $(\{3 \omega_4+5 \omega_3+4\Delta_4\}^2,$\\
$\#_2(u_{l+1})=4$& $\omega_4+4\omega_3)$ &$ \{2\omega_4+5\omega_3+4\Delta_4\}^2)$& $ \{2\omega_4+5\omega_3+4\Delta_4\}^2)$\\
upper bounds &\rut{1.1136^K} &\rut{1.1138^K} & \rut{1.1143^K}\\ \hline 
$\#_2(u_0)=4$&$(\{4\omega_4+4\omega_3+4\Delta_4\}^2,$ &case $\beta$ instead &$(\{4\omega_4+4\omega_3+4\Delta_4\}^2,$\\
$\#_2(u_{l+1})=4$&$,\omega_4+4\omega_3)$&  &$\{2\omega_4+4\omega_3+4\Delta_4\}^2)$\\ 
upper bounds & \rut{1.1088^K}&  & \rut{1.1088^K}\\ \hline 
\end{tabular}\\[1ex]
\caption{}
\label{tabb1}
\end{table}\\
We have a $(\{2\omega_4+7\omega_3+4\Delta_4\}^2,\omega_4+4\omega_3)$-branch for case $\alpha)$ which \rut{1.1132^K} properly upper bounds. We also have a  $(\omega_4+4\omega_3+2\Delta_4,\{2\omega_4+5\omega_3+4\Delta_4\}^2)$-branch for case $\beta)$ such that it is upper bounded by \rut{1.1142^K}.

\end{appendix}
\end{document}